\documentclass[useAMS,usenatbib]{mn2e}
\usepackage{graphicx}

\title[The Fundamental Manifold of spiral galaxies]{The Fundamental Manifold of spiral galaxies: ordered versus random motions and the morphology dependence of the Tully-Fisher relation.}

\author[C. Tonini et al.]
{C. Tonini$^{1}$
\thanks{E-mail:ctonini@astro.swin.edu.au},
D. H. Jones$^{2}$,
J. Mould$^{1}$,
R. L. Webster$^{3}$, 
T. Danilovich$^{4}$,
and S. Ozbilgen$^{3}$ \\  
\\
$^{1}$Centre for Astrophysics and Supercomputing, Swinburne University
of Technology, Hawthorn, VIC 3122, Australia\\
$^{2}$School of Physics, Monash University, Clayton, VIC 3800, Australia\\
$^{3}$School of Physics, University of Melbourne, Parkville, 3010 VIC, Australia\\
$^{4}$Chalmers University, Maskingrand 2, 412 58 Goteborg, Sweden \\
}

\begin{document}

  

\maketitle

\begin{abstract}

We investigate the morphology dependence of the Tully-Fisher (TF) relation, and the expansion of the relation into a three-dimensional manifold defined by luminosity, total circular velocity and a third dynamical parameter, to fully characterise spiral galaxies across all morphological types.
We use a full semi-analytic hierarchical model (based on Croton et al. 2006), built on cosmological simulations of structure formation, to model galaxy evolution and build the theoretical Tully-Fisher relation. With this tool, we analyse a unique dataset of galaxies for which we cross-match luminosity with total circular velocity and central velocity dispersion. We provide a theoretical framework to calculate such measurable quantities from hierarchical semi-analytic models. 
We establish the morphology dependence of the TF relation in both model and data. We analyse the dynamical properties of the model galaxies and determine that the parameter $\sigma/V_{\rm C}$, i.e. the ratio between random and total motions defined by velocity dispersion and circular velocity, accurately characterises the varying slope of the TF relation for different model galaxy types. 
We apply these dynamical cuts to the observed galaxies and find indeed that such selection produces a differential slope of the TF relation. 
The TF slope in different ranges of $\sigma/V_{\rm C}$ is consistent with that for the traditional photometric classification in Sa, Sb, Sc. We conclude that $\sigma/V_{\rm C}$ is a good parameter to classify galaxy type, and we argue that such classification based on dynamics more closely mirrors the physical properties of the observed galaxies, compared to visual (photometric) classification. We also argue that dynamical classification is useful for samples where eye inspection is not reliable or impractical. 
We conclude that $\sigma/V_{\rm C}$ is a suitable parameter to characterise the hierarchical assembly history that determines the disk-to-bulge ratio, and to expand the TF relation into a three-dimensional manifold, defined by luminosity, circular velocity and $\sigma/V_{\rm C}$. 

\end{abstract}


\section{Introduction}
 
Galaxy scaling relations highlight the regularities that characterise galaxy formation. Of particular interest is the physical connection between the dynamical state of a galaxy, determined by its assembly history, and the galaxy photometric properties, tracing its stellar populations and its star formation history. 
The main dynamical quantity in such an analysis is the galaxy total gravitational potential, which determines the galaxy rotation curve. In the case of spiral galaxies, this is usually parameterised with the circular velocity at a given radius, and in the case of ellipticals, with the central velocity dispersion. Both these quantities have a monothonic dependence on luminosity, thus giving rise to the Tully-Fisher relation (TF; Tully $\&$ Fisher, 1977) for spirals, and the Faber-Jackson relation (FJ; Faber $\&$ Jackson 1976) for ellipticals.

However galaxy structure cannot be completely captured by such simple scalings, and the underlying complexity emerges with additional parameters. In the case of ellipticals, the triaxial mass distribution causes an expansion of the FJ relation into a third dimension, recasting it as a scaling between velocity dispersion $\sigma$, surface brightness $I_e$ and effective radius $r_e$, known as the Fundamental Plane (FP; Djorgovski $\&$ Davis 1987, Dressler et al. 1987), a manifestation of the virial theorem in observational quantities (see for instance Cappellari et al. 2006). The slope of the FP indicates that the structure of elliptical galaxies is not a self-similar scaled version of a single object as a function of mass. 

In the case of spirals, the apparent simplicity of these systems is broken by the observational evidence of a morphology dependence of the TF slope (Springob et al. 2007, Masters et al. 2008). Such feature is also predicted by hierarchical galaxy formation models (Tonini et al. 2011), which naturally produce a varying TF slope that depends on the galaxy internal dynamics, a signature of its assembly history. Indeed, there is an extensive body of work in the literature supporting the observational evidence of a dependence of the TF on galaxy type; the TF for late-type spirals (Masters et al. 2006, Courteau 1997, Giovanelli et al. 1997) is not followed by early-type spirals and S0 types (Williams et al. 2010, Bedregal et al. 2006, Neistein et al. 1999), dwarf galaxies (McGaugh et al. 2000, Begum et al. 2008), barred spirals (Courteau et al. 2003), and polar ring galaxies (Iodice et al. 2003). 

The evidence suggests that a third parameter beside circular velocity and luminosity enters the TF relation. This parameter depends on galaxy type, and expands the TF into a three-dimensional manifold that describes the structure of spirals or alternatively, of all galaxies. 
A general three-dimensional manifold for galaxies regardless of type has been investigated (see Zaritsky et al. 2008), with the set of observables ($V_c, I_e, r_e$) in analogy with the ellipticals FP. The approach of this work is the use of analytic toy models, which cannot capture the intrinsic complexity of galaxies due to their hierarchical mass assembly, but resort to absorb it into a single parameter, $\Gamma_e$, the mass-to-light ratio inside the half-light radius $r_e$, and to fit this parameter assuming that galaxies lie on the manifold. Catinella et al. (2012) make use instead of dynamical indicators such as rotational velocity and dispersion to obtain a generalised baryonic FJ relation that holds for all galaxy types. Courteau et al. (2007) investigate galaxy size, and analyse the spiral size-luminosity and luminosity-velocity relations and their dependence on morphology and stellar population content. 

In this work we consider two questions: 1) is there a third dynamical parameter that characterises the morphology-dependence of the Tully-Fisher relation, and can be used to expand such relation into a three-dimensional manifold, describing the structure of spiral galaxies?; and 2) can we use this parameter to classify galaxy morphology, when visualisation is not available (for instance at high redshift) or impractical (for instance for large surveys)?

The novelty of our analysis is that we take advantage of a full semi-analytic hierarchical model (based from Croton et al. 2006)
to define a dynamical parameter that characterises galaxy morphology, predicts the TF relation for different galaxy types and defines the spiral galaxy manifold.
The model is built on cosmological simulations of structure formation, to model galaxy formation and evolution and build the theoretical TF relation. This tool is ideal for accounting for the merger history of galaxies and their complex star formation history, which is recorded in their rotation curve and stellar populations. We take particular care in producing dynamical and photometric quantities that can be directly compared with observations. 
With this tool, we analyse a unique dataset of galaxies: we build a sample of observed galaxies carrying the information on luminosity, circular velocity and central velocity dispersion, derived from the GALEX Arecibo SDSS Survey (Catinella et al. 2013). To this sample we apply our theoretical predictions.

In Section 2 we describe the model. In Section 3 we introduce our observational sample and present our main results. In Section 4 we discuss our findings and present our conclusions.

\section{The model}     

Semi-analytic models are a powerful tool for investigating the TF relation in a cosmological framework. They naturally interlink the dynamics of structure formation with the galaxy emission; the galaxy assembly and star formation histories derive directly from the hierarchical growth of structures. In addition, such models allow for a thorough statistical analysis. 
The model galaxies are obtained with the semi-analytic model by Croton et al. (2006), with the spectro-photometric model (including dust absorption and emission) described in Tonini et al. (2012). We implement the prescription for the galaxy rotation curves by Tonini et al. (2011), where the velocity profile is determined by the mass distribution of all galaxy components (dark matter, stellar disk and bulge, and gas), and the total circular velocity is 
\begin{equation}
V_{\rm C}^2(r)=V^2_{\rm DM}(r)+V^2_{\rm disk}(r)+V^2_{\rm bulge}(r) 
\label{Vc}
\end{equation}
Each of the velocity terms in the equation is of the type $V^2 \propto G \ M(r)/r$ where $r$ is the galactocentric radius and $M(r)$ is the mass profile: a truncated isothermal sphere for the dark matter halo, a flat exponential disk, and a Hernquist (1990) profile for the bulge (see Tonini et al. 2011 for a complete description; see also Tonini et al. 2006a, Salucci et al. 2007).
The slope of the TF relation in models, and in particular its tilting with galaxy type observed in the data, is the manifestation of the connection between galaxy dynamics and star formation and assembly history. For spiral galaxies in general, morphology can be understood in terms of bulge-to-disk ratios. Dynamically, the growth of a bulge in the center of a disk is the result of secular (angular momentum redistribution and stellar migration, bars) and violent (mergers) processes, all of which leave a trace in the galaxy rotation curve. At the same time, the star formation history of the galaxy, which is affected by the dynamical evolution, imprints the bulge-to-disk luminosity. A theoretical determination of the TF relation needs to incorporate each of these effects to be successful, a job that hierarchical semi-analytic models are best suited to accomplish. 

\subsection{Which radius? An angular momentum problem}

When studying the morphology dependence of the TF relation, it becomes especially important to measure the TF at a meaningful, physically motivated radius. Because of the different radial profiles of the galaxy dynamical components (disk, bulge, dark matter halo, gas), \textit{1)} the slope of the TF relation varies with the radius at which the rotation velocity is calculated (Yegorova et al. 2007) and \textit{2)} the same radius in galaxies of different morphology probes different dynamical regions, thus introducing an artificial scatter in the TF. Traditionally, the velocity at a galactocentric radius $r=2.2R_{\rm D}$, where $R_{\rm D}$ is the exponential stellar disk scale-length, has often been used to build the TF relation from observations. This velocity roughly corresponds to the peak velocity of a bulgeless disk; in the case of a galaxy with a substantial bulge however, the region around $r=2.2R_{\rm D}$ contains a different mix of dark matter, gas and stars, and the peak of the rotation curve is actually at a different radius.

Following Tonini et al. (2011), we adopt a \textit{dynamical} definition of the disk scale-length, that corresponds to a \textit{fixed angular momentum} rather than a fixed galactocentric distance. With this definition the value of $R_{\rm D}$ self-regulates in the presence of a bulge. The formation of the bulge, both from secular evolution and mergers, implies that stars migrate radially or with inspiralling orbits to the centre of the galaxy, losing all their angular momentum and settling into a pressure-supported configuration.
This lost angular momentum is transferred to the disk (Dutton et al. 2007; see also Tonini et al. 2006), with the net effect of increasing the disk size. For a bulge of mass $M_{\rm bulge}$ forming in a disk of mass $M_{\rm disk}$ with initial scale-length $R_{\rm D_{\rm old}}$, the disk scale-length after angular momentum transfer is
\begin{equation}
R_{\rm D}=R_{\rm D_{\rm old}} \left( 1+(1-f_{\rm x}) \frac{M_{\rm bulge}}{M_{\rm disk}} \right)~,
\label{rdafter}
\end{equation}
with the fiducial value $f_{\rm x}=0.25$ indicated by Dutton et al. (2007). The new $R_{\rm D}$ represents a 'corrected' disk scale-length, that takes into account the additional gravitational potential of the bulge. After the correction, all galaxies move onto the disk mass-disk scale length relation that holds for Sc galaxies (see Tonini et al. 2011). 
After definining a $R_{\rm D}$ that evolves with morphology, it then makes physical sense to adopt $r=2.2R_{\rm D}$ as our radius of choice to build the TF relation. 

\subsection{Integrated galaxy velocity dispersion}

From the mass distribution we can build velocity profiles for all galaxy components. The theoretical velocity dispersion is the sum of the contribution to the rotation curve by all components that are pressure-supported, namely the dark matter halo and the bulge:
\begin{equation} 
\sigma(r)=\sqrt{V^2_{\rm DM}(r)+V^2_{\rm bulge}(r)}~,
\label{sigma}
\end{equation}
where each velocity term in the equation is determined by the mass profile of the dynamical component: $V_i^2(r) \propto GM_i(r)/r$ (see Tonini et al. 2011 for a detailed description).
However this particular kind of output is not readily comparable with observations. In fact, in the literature the observed samples usually provide a single-value galaxy velocity dispersion for each object, which is obtained from the broadening of distinct spectral features due to the internal motions of the stars. This measure is an integrated quantity over a galactocentric radius generally determined by telescope aperture or detection limit. The model on the other hand outputs intrinsic, physical galaxy properties, thus the comparison with the observed spectral line dispersion requires \textit{1)} the definition of a 'model aperture radius' inside which to compute the velocity dispersion, and \textit{2)} the definition of an integrated velocity dispersion inside this radius.  

In defining such a radius, there are two factors to consider: \textit{1)} the galaxy dynamics becomes increasingly dark matter-dominated at larger galactocentric distances, but \textit{2)} the only visible tracers of the galaxy velocity dispersion are the stars in the bulge, since the disk is modeled as completely rotation-supported. Therefore, if we were to 'observe' a model galaxy, the entirety of the velocity dispersion signal in the spectral features would come from the stars in the bulge. For this reason, we assume that the 'model aperture radius' corresponds to the bulge outer limit. The latter is calculated as $R_{out}=3.5 R_S$, where $R_S$ is the characteristic scale-length of the Hernquist density profile (following Tonini et al. 2011; this is consistent with other semi-analytic models, for instance GalICS, Hatton et al. 2003). $R_S$ is not well constrained from observations and its relation to other physical parameters is quite uncertain, depending on the formation mechanisms of bulges (for instance, are bulges formed in merger events, or are they formed through secular evolution from the disk?). For this reason, we introduce a random scatter in the value $R_S$, only costrained to be at most half of the disk scale-length $R_D$, which is reasonable for spiral galaxies. 

The width of a spectral line is in principle obtained by averaging the velocity dispersion over all the stars; in models, that calculate theoretical velocity dispersion profiles, the equivalent quantity is the mass-averaged velocity dispersion, calculated over radial bins out to the bulge outer limit: 
\begin{equation}
\sigma = \frac{\sum_n M_n \ \sigma_n}{\sum_n M_n}~,
\label{massweight}
\end{equation}
where $M_n$ is the bulge mass in the $nth$ shell (out to the bulge outer limit), and $\sigma_n$ is the total velocity dispersion in the centre of the bin, calculated from Eq.~(\ref{sigma}). 

\subsection{Selection of the model galaxies}

In the model, morphology can be defined in terms of the galaxy physical parameters, like the mass of the bulge and the disk (see Tonini et al. 2011). This method has the advantage of grouping together objects that share a similar formation history, thus favouring a more detailed study of the physics involved in their evolution. On the other hand, this type of selection is hard to apply to observations, and it involves some model-dependencies in the conversion between colors and luminosities to masses and ages, thus confusing
the comparison between models and data.

\begin{figure}
\includegraphics[scale=0.3]{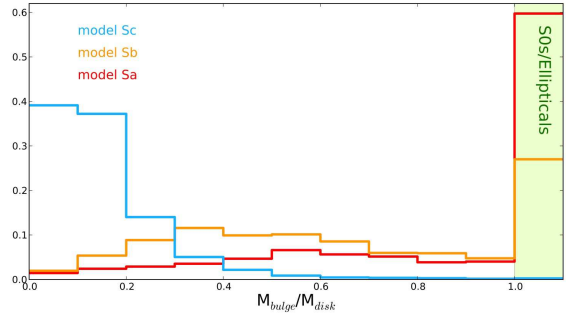}
\caption{The distribution of the mass ratio $M_{\rm bulge}/M_{\rm disk}$ for the model Sc, Sb and Sa galaxies (\textit{blue, orange and red respectively}).} 
\label{histo}
\end{figure}

In most observations the determination of the Hubble type employs the use of some photometric criterion, based on colors or on the relative luminosities of bulge and disk when available. To facilitate the comparison, we classify our model galaxies with one of such methods, based on the bulge-to-total luminosity ratio in the B band, following Simien $\&$ De Vaucouleurs (1986): after defining $\mu_B=M_B(bulge)-M_B(total)$,  Sc galaxies are characterised by $2.3 < \mu_B < 4.15$, Sb galaxies by $1.23 < \mu_B < 2.01$ and Sa galaxies by $0.8 < \mu_B < 1.23$. In Fig.~(\ref{histo}) we compare this classification with the actual mass ratio of bulge and disk, $M_{bulge}/M_{disk}$. With this classification, we find that the Sc types are well represented by galaxies with small bulges (less than $\sim 20 \% $ of the disk mass). On the other hand, Sb types show a wide variety of bulge-to-disk ratios (peaking between 0.2 and 0.7), and Sa galaxies are rare below 1, where they show a very flat distribution. Both Sb and Sa types leak into the S0 and elliptical regimes, defined in the model for values $M_{bulge}/M_{disk}>1$. It is not difficult to imagine that such a scenario is present in observed galaxies too, especially if a band-pass is used where the luminosity evolves rapidly with the age of the stellar populations, and is moreover subject to dust extinction, like the B band.
The photometric definition of Sb and Sa model galaxies select objects that do not belong to a uniform population in terms of physical parameters and formation history. 

In general photometric classifications may rely on more than one band, and the accuracy of the classification increases with the number of bands available. However, the photometric classification relies on the fact that the stellar populations of bulge and disk are \textit{visibly} different, and that might not be the case, especially for more massive and evolved systems. Even the most sophisticated photometric schemes in fact cannot easily disentangle the effects of age, metallicity, and reddening (see Pforr et al. 2012, 2013).
The degeneracies in luminosity and colours of different stellar populations confuse the mass-to-light ratio and the decomposition of bulge and disk, and we believe this is one factor at the origin of the increased observed scatter in the TF relation for these galaxy types. In fact, the mass-to-light ratio, by mirroring the balance between different dynamical components, is the main factor that shapes the rotation curve, which in turn determines the slope of the TF relation.

\subsection{Comparison with observations}

To compare the model with observed data, we employ the sample of massive galaxies of GASS (GALEX Arecibo SDSS) survey, Data Release 3 (Catinella et al. (2013; see also Catinella et al. 2010 for a complete description of the survey), which provides circular velocities obtained from HI linewidths. To this sample we add measurements of the central velocity dispersions measurements from the Sloan Digital Sky Survey, Data Release 9 (Ahn et al. 2012), and K-band magnitudes from the Two Micron All Sky Survey (2MASS; Skrutskie et al. 2006). Distances are estimated from the 21 cm redshifts (see Mould et al 2000). To minimise the scatter we consider only galaxies with inclinations $> 45^{\circ}$. The final sample after cutting away non-detections and poor-quality detections (see Catinella et al. 2013) consists of $\sim 340$ galaxies.

The comparison between model velocities and observed quantities is complicated by some unavoidable caveats, that we address below. 

$\bullet$ In the GASS sample the circular velocity is obtained from HI linewidths. The HI linewidth is an integrated quantity, that we compare with the amplitude of the theoretical rotation curve at one radius. However $r=2.2 R_D$ is large enough for the rotation curve to have already peaked; following Catinella et al. (2006) and Verheijen (2001), we know that in massive galaxies (such as those in the GASS sample) the rotation curve rises rapidly and it remains flat for most of the radial range. Therefore, a measurement of the HI linewidth will be largely dominated by the signal coming from gas that would contribute to the flat region of the rotation curve. Being this the case, the HI linewidth will yield a value of the circular velocity consistent with that which would be measured from the rotation curve itself.

$\bullet$ A source of scatter in the Tully-Fisher, whether theoretical or observational, comes from the choice of a radius at which to measure it, or the choice of an aperture within which to collect the signal and determine a linewidth. As described in Yegorova et al. (2007) and Salucci et al. (2007), rotation curves are actually rarely really flat, and measurements encompassing different radii will give rise to different slopes of the TF relation. This
is due to the different density profiles of the galaxy components (disk, bulge, dark matter halo), each of which dominate the curve at different radii. As different radii map different dynamical regimes in the galaxy, so the signal in the HI linewidth, or the amplitude of the rotation curve, will be dominated by different components depending on aperture or radius. In the model we choose a radius large enough to be well away from the bulge, and moreover apply our correction of $R_D$ in Eq.~(\ref{rdafter}) that 'resets' the curve in the presence of the bulge. However the farther out we go, the more the curve is dominated by the dark matter halo. As already pointed out, for massive spiral galaxies such as those in the GASS sample the shape of the rotation curve is well behaved outside the bulge (Catinella et al. 2006, Verheijen 2001). However in general, as described by Verheijen (2001), the shape of the curve depends on the galaxy mass and type, and in particular the radius at which the dark matter halo starts to dominate the curve is smaller and smaller for less massive galaxies, with dwarfs having a constantly rising curve out to the last measured point. Therefore, an important source of scatter in the TF relation is driven by the shape of the curve, $dV/dr$, which in turn is determined by the baryonic/dark matter mass ratio or the total $M/L$ ratio (as has been highlighted for the B band in Fig.~\ref{histo}). 
 
$\bullet$ In the model the velocity dispersion is traced by the bulge stars, since the disk is fully rotationally supported, and therefore we 
set our "aperture'' as encompassing the radial extension of the bulge. Note that this implies a different radius for each galaxy, and corresponds to the assumption that the aperture is always larger than the target. The velocity dispersion for the observed sample on the contrary is determined by the SDSS fiber line-width, which has a fixed aperture centered on the centre of each galaxy. We argue that this difference is however smaller than the scatter we obtain in the theoretical values of the velocity dispersion $\sigma$, dominated by the intrinsic scatter in the galaxy assembly histories.

$\bullet$ Regarding the theoretical velocity dispersion, a quantity more directly comparable with data would be a luminosity-averaged (rather than mass-averaged) dispersion, but this would require an additional layer of modeling, in particular of the radial dependence of the mass-to-light ratio, that is not very well constrained observationally and represents an important source of scatter. In effect our implementation corresponds to assuming a constand mass-to-light ratio for the bulge stars; although this is not strictly true in general, it is nonetheless a fairly reasonable assumtpion for the K band, given that the bulge stars are mostly old (a few Gyrs) and that the spatial extension of the bulge is rather small compared to the disk. In addition, any differences between the mass profile and the K-band luminosity profile are smoothed out by the averaging operation (checked for convergence over different binning grids), and only very large differences, which are unlikely in the K band, would produce significantly different values of $\sigma$. 

$\bullet$ We do not include the disk contribution to $\sigma$. The disk is modeled as fully rotationally supported, so the disk stars are dispersion-free. In addition, we find that the disk contribution to the total gravitational potential inside the bulge radius is negligible and its effect on the calculated $\sigma$ is marginal. Observationally the inner disk would add to the signal coming from the bulge, depending on inclination. The model galaxies do not suffer from inclination effects, and rather than attempting to model them, we choose to use data that have been corrected for inclination. In addition we argue that, since we are restricting ourselves to the K band, the disk contribution to the estimate of $\sigma$ in observed galaxies is not likely to be significant.

\section{Results}

The observational sample we use is not a Tully-Fisher dedicated sample, and galaxies were not selected based on the quality of their rotation curves or their morphological type. The goal in this work is to characterise the physical properties of the galaxies in this sample so that they would naturally produce a morphology-dependent TF relation. In other words, we want to find the physical parameters that uniquely characterise the Tully-Fisher manifold of spiral galaxies, by which a galaxy type is defined based on dynamics and physical parameters linked to its assembly history, rather than visual inspection.

\begin{figure}
\includegraphics[scale=0.45]{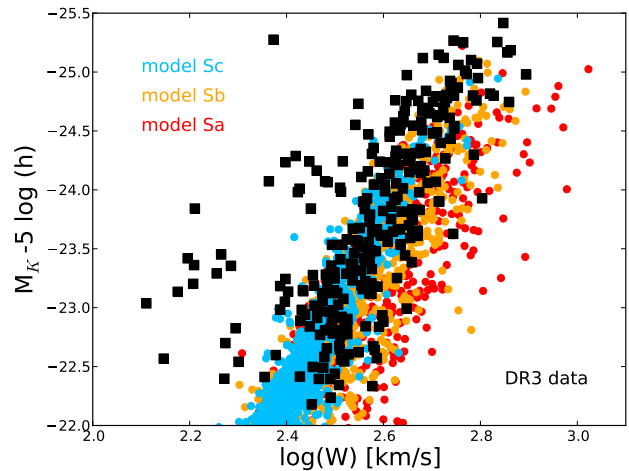}  
\caption{The Tully-Fisher relation for the semi-analytic model galaxies, divided according to morphology (\textit{light blue, orange and red points} for Sc, Sb and Sa galaxies respectively), compared with the observed sample (\textit{black squares}).}
\label{tfmodel}
\end{figure}

We start by building the TF relation for the observed sample, plotted in Fig.~(\ref{tfmodel}) in terms of $W=2 V_{\rm C}$ and represented by the \textit{black squares}. Note that the majority of the galaxies in the sample follow a reasonably-defined relation, but with a rather large scatter. In addition, there are numerous objects that significantly deviate from such behaviour. Fig.~(\ref{tfmodel}) also shows the Tully-Fisher relation for the model galaxies, divided according to morphology: \textit{light blue, orange and red points} represent Sc, Sb and Sa galaxies respectively. 
As the bulge-to-disk mass ratio increases, the slope of the TF relation flattens, due to the combination of two effects: 1) the stellar populations in the bulge are on average older and therefore their $M/L$ is higher (for instance in the K band, the very luminous post-main sequence phases have faded away; Maraston 2005), while at the same time the velocity is higher due to the boost the bulge inflicts on the rotation curve, and 2) the scatter in the bulge-to-disk mass ratio increases with mass, because larger galaxies have a wider variety of merger histories. Therefore an Sa galaxy that by chance has a smaller than average bulge will scatter back towards the main TF relation, while an Sa galaxy with a larger than average bulge will scatter further away from the TF, and more so with increasind galaxy mass. The overall effect is that of flattening the TF slope. In addition, at the very high-mass end of the mass distribution AGN feedback causes a further increase of the $M/L$ ratio. 

Model and data roughly occupy the same locus in the plot (ignoring the data outliers for the moment), so we use the model to provide insight into the observed galaxies.

In the model, the galaxy type and its evolutionary history (short $vs$ prolongued star formation history, merger-rich $vs$ merger-poor assembly) is at first order characterised by the mass ratio between the spherical and disky components $M_{\rm bukge}/M_{\rm disk}$. A more observationally-friendly quantity to express this is the ratio between random and total motions. In the model this corresponds to
$\sigma/V_{\rm C}$, where $\sigma$ is defined by Eq.~(\ref{sigma}), and $V_{\rm C}^2=\sigma^2+V_{\rm disk}^2$ (Eq.~(\ref{Vc})).
The bulge and dark matter halo mostly contribute with velocity dispersion, and the disk mostly with rotation. 

\begin{figure*}
\includegraphics[scale=0.75]{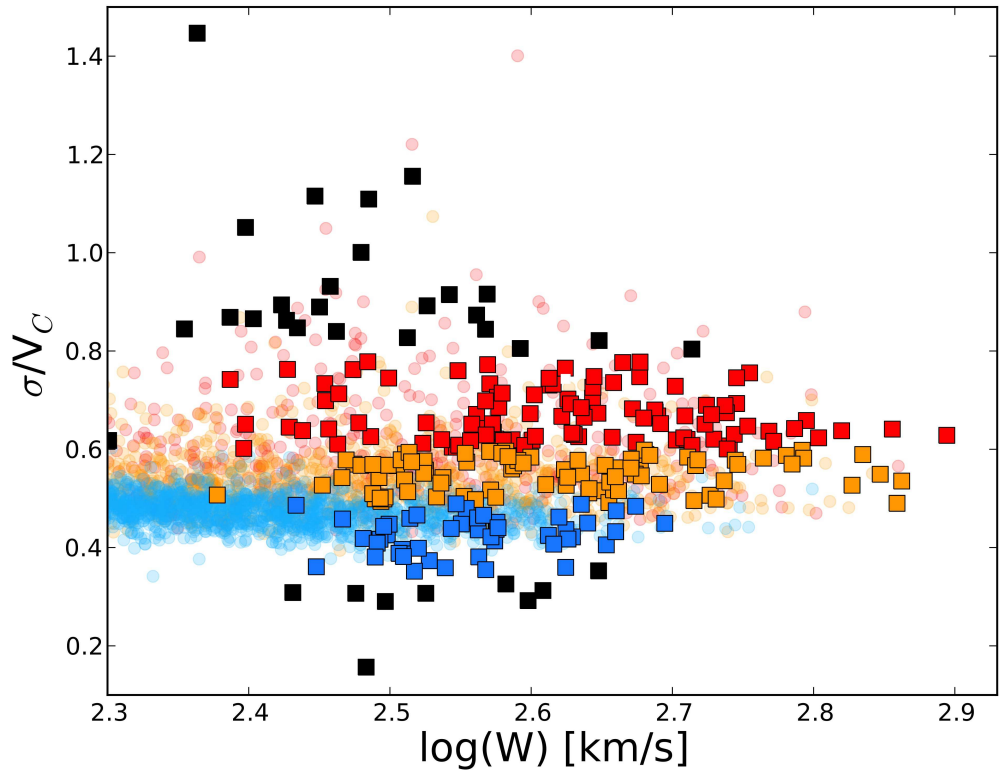}   
\caption{The ratio of velocity dispersion and circular velocity as a function of $W=2 V_{\rm C}$, for the model galaxies (\textit{open circles, colour-coded as in Fig.~(\ref{tfmodel})}) and for the observed sample (\textit{squares, colour coded depending on the locus they share with the model}). Data outliers are represented in \textit{black}. } 
\label{sigmamodel}
\end{figure*}

After splitting the model galaxies in photometric classes Sa, Sb and Sc, we consider the ratio $\sigma/V_{\rm C}$ for each type, and plot it in Fig.~(\ref{sigmamodel}), as a function of the total circular velocity $V_{\rm C}$, calculated at $2.2 \ R_{\rm D}$ (where $R_{\rm D}$ is defined by Eq.~(\ref{rdafter})); these are the \textit{circles, colour-coded as in Fig.~(\ref{tfmodel})}. 
The model galaxies show a differentiation in $\sigma/V_{\rm C}$ depending on the B-band selected morphology, with later-type spirals showing a smaller total velocity dispersion than the earlier types, for a given total circular velocity, and a smaller scatter. This shows how dynamics and star formation history are interlinked in the model: the hierarchical build-up of galaxies grows bulges through mergers and evolves the star formation rates to produce redder early-type objects. Moreover, following the nature of hierarchical mass assembly, earlier-type galaxies live in halos with a richer and more varied merger history, that affects both the bulge and the halo mass; this increases the scatter in $\sigma/V_{\rm C}$.
Notice, however, how $\sigma/V_{\rm C}$ does not depend on the total circular velocity $W$, for a given morphological type. 

We then compare the observed sample in the same space (\textit{squares}). The morphology of the galaxies in the sample is not known, and we use the relation between photometric type and $\sigma/V_{\rm C}$ of the model to classify them: we colour-code the data based on where they lay with respect to the model, in bands of $\sigma/V_{\rm C}$. We assign \textit{blue} colour to the galaxies with very low velocity dispersion, with values of $\sigma/V_{\rm C}$ typical of model Sc galaxies, \textit{red} to the more dispersion-dominated, with values of $\sigma/V_{\rm C}$ typical of model Sa, and \textit{orange} to the intermediate class, typical of model Sb objects.
There is a certain degree of overlap in the model, i.e. there are ranges of $\sigma/V_{\rm C}$ that are occupied by both model Sb and Sa galaxies and, to a lesser degree, by both model Sc and Sb (in analogy with Fig.~(\ref{histo})).
We consider the $\sigma/V_{\rm C}$ intervals with a clear predominance of one type above the others, and colour-code the data accordingly, keeping in mind that the overlap between classes is going to be a source of scatter (particularly between \textit{orange} and \textit{red} data points). 

In addition, we colour in \textit{green} the outliers, i.e. galaxies that do not fall in the locus of the model (those with $\sigma/V_{\rm C} > 0.8$ and $\sigma/V_{\rm C} < 0.35$). We also consider as outliers all objects with $V_{\rm C}<100 \ km/s$, which fall way off both the bulk of the data and the model in the TF plot (Fig.~(\ref{tfmodel}); moreover, the model starts to suffer from resolution effects at those masses, given the mass resolution of the Millennium simulation). 

The outliers are mostly dispersion-dominated (in accord with the analysis by Catinella et al. 2012), a feature that in the model is a signature of an early-type/elliptical galaxy. A few outliers show instead a very low velocity dispersion. However, in the range of circular velocities considered here, ratios $\sigma/V_{\rm C}<0.35$ yield velocity dispersions $\sigma < 70 \ km/s$, which is the resolution limit of the Sloan spectrograph. For this reason, such values of the velocity dispersion cannot be considered reliable (Bernardi et al. 2003).

Is $\sigma/V_{\rm C}$ a good parameter to characterise observed galaxies? In other words, have we selected a spiral sample, based entirely on theoretical expectations of the ratio between random and total motions? And is this classification a good proxy for galaxy morphology, i.e. will the sub-classes produce different TF relations, in accord with observations?

\begin{figure*}
\includegraphics[scale=0.75]{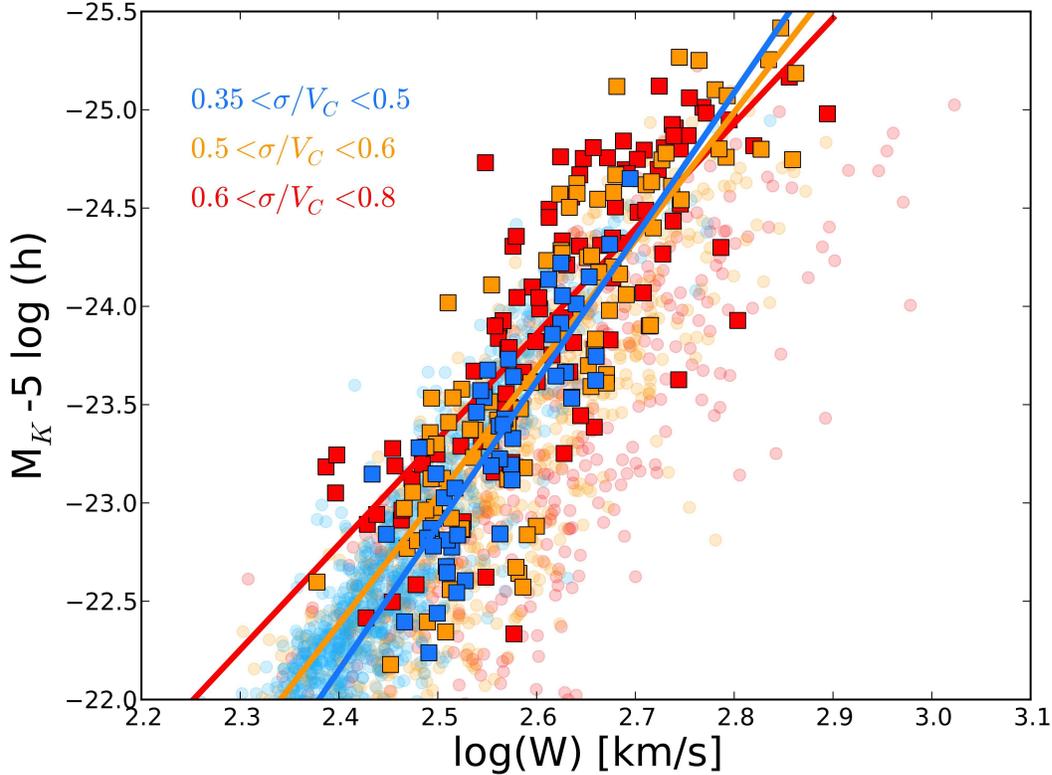}   
\caption{The TF relation for the observed sample, after splitting it into morphological types based on the model $\sigma/V_{\rm C}$, as in Fig.~(\ref{sigmamodel}). The lines represent linear regression fits. The \textit{circles} represent the model galaxies classified as Sa, Sb and Sc. The colour-coding is the same as in Fig.~(\ref{sigmamodel}). }
\label{tfmodel2}
\end{figure*}

\begin{figure*}
\includegraphics[scale=0.35]{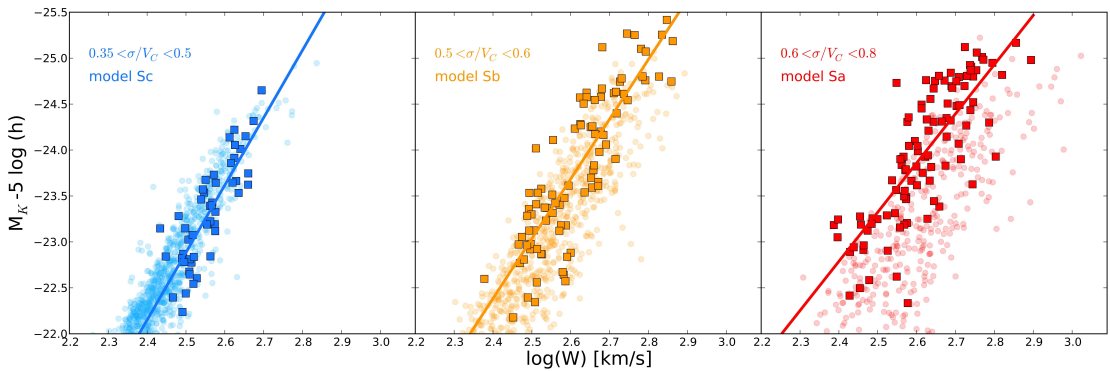}   
\caption{The TF relation for the observed sample, same as in Fig.~(\ref{tfmodel2}), but here we split the TF into 3 panes, one for each interval of $\sigma/V_{\rm C}$. The lines represent linear regression fits. The \textit{circles} represent the model galaxies classified as Sa, Sb and Sc (\textit{left to right}). Colour-coding is as in the previous figures.}
\label{tfmodel3}
\end{figure*}

Fig.~(\ref{tfmodel2}) shows again the TF relation for the observed sample, after we have divided the galaxies in bands of $\sigma/V_{\rm C}$, following the prediction of the semi-analytic model described in Fig.~(\ref{sigmamodel}). In the same plot we show the model galaxies (\textit{circles}). In addition in Fig.~(\ref{tfmodel3}) we show the same plot, split into 3 panels for clarity. 

Indeed, the three subsamples follow three distinct TF relations. As bulges grow larger in galaxies, the TF slope flattens as expected, mostly due to the increased mass-to-light ratio due to the spherical component. At the same time, the scatter increases; as this happens both in the data and in the model, this must be intrinsic rather than due to increased uncertainties in the measurements of $W$ and $M_K$. In models, the increased scatter is due to the hierarchical nature of galaxy assembly, more prominent in bulge-heavy galaxies because the bulge is a signature of a significant merger history (Tonini et al. 2011; see also Tonini, 2013). 

A linear regression\footnote{Least Squares fit with two coefficients: slope and zero-point.} of the TF relation for the three subsamples yields the values $[-5.35 \pm 0.40, -6.50 \pm 0.42, -7.34 \pm 0.72]$ for the TF slope of Sa, Sb and Sc-types respectively (also consistent with the theoretical values obtained in Tonini et al. 2011, and the observations of Masters et al. 2008). The best fit lines are shown in Figs.~(\ref{tfmodel2}, \ref{tfmodel3}) in \textit{red, orange and light blue} colours respectively. The model galaxies are shown in the background, colour-coded as usual. 

The slopes and zero-points for the more rotationally-supported objects (\textit{light blue and orange lines}) are well matched with the model Sc and Sb TF relations. The TF slope for the observed more dispersion-dominated objects is consistent with that of model Sa galaxies, although the zero-point is offset by about $~0.5 \ mag$ or by $~50 \%$ in the velocity. This might be due to selection effects. In fact, the GASS sample was selected based on the HI signature; while in the model Sc and Sb galaxies are pretty uniformly gas-rich, the earlier types present a larger scatter in gas content, depending on the assembly history, and therefore this class of model objects cannot be matched in its entirety by the GASS sample. 

As a sanity check, we also visually inspect the observed galaxy sample, to verify to what degree the morphology selection based on $\sigma/V_{\rm C}$ mirrors the traditional Sa, Sb and Sc classification. We find that for about $70 \%$ of the sample the two selections match exactly, with an expecially good match for later types, while for the remaining galaxies there is an offset of one type, predominantly involving Sb-type objects. Notice how this is in agreement with the predicted uncertainty regarding Sb galaxies pointed out in Fig.~(\ref{histo}).

Fig.~(\ref{tfmodel2}) shows that $\sigma/V_{\rm C}$ is a good dynamical parameter to characterise spiral galaxies, and it is a good proxy for morphology. The data selected based on $\sigma/V_{\rm C}$ produces morpholgy-dependent TF relations, consistent with the theoretical expectations and previous observational results. The spiral galaxy scaling relation between luminosity and dynamics seems fully characterised in the three-dimensional space $[M_{\rm K}, V_{\rm C}, \sigma/V_{\rm C}]$ across all spiral types.

\section{Discussion and conclusions}

The Tully-Fisher relation is the product of virial equilibrium combined with the star formation history. It links a measure of the total gravitational potential, $V_{\rm C}$, with the luminosity in various bands, which maps the stellar mass as a function of age. The fact that the TF relation holds for all spirals shows that there is one principal parameter that largely governs galaxy evolution, i.e. the total mass. On the other hand, the perturbations in the TF, such as the observed morphology dependence, indicate that at least a second parameter plays a detectable role. Such parameter is linked to the mass \textit{distribution} inside the galaxy, a product of both secular evolution and hierarchical mass assembly (an ideal scenario to study with a semi-analytic model). 

The mass distribution might be parameterised with some definition of effective radius, as in the case or ellipticals, but this method relies on a photometric classification that in both models and observations is affected by systematics. Another route is to consider that mass distribution is interlinked with the distribution of the internal motions. In these terms, a dynamical parameter such as $\sigma/V_{\rm C}$ represents as well the mass concentration as the fraction of random over total motions inside the galaxy. In the formalism of virial equilibrium phase-space analysis, it represents an angular momentum parameter. This makes it a very clear-cut, physically well defined quantity to determine in models, but it also has observational advantages. 
In fact, while a morphology analysis or the determination of effective radii is very uncertain for distant galaxies, central velocity dispersion and circular velocity are relatively easy to determine from galaxy spectra, pushing the analysis of the TF relation to higher redshifts. In addition, this method is ideal in the case of large galaxy surveys, where a classification based on visual inspection is impractical.

The parameter $\sigma/V_{\rm C}$ is tracked by the emission of a fraction of the stellar populations, those that are not rotationally supported. These are mostly the bulge stars, a component of intermediate to old age, with a higher mass-to-light ratio than the disk. The prominence of this component causes the velocity-luminosity relation to shift from that of the Sc types (close to pure disks). Thus $\sigma/V_{\rm C}$ is a good proxy for galaxy morphology, it directly relates to the bulge-to-total mass ratio, which in turn can be linked to the bulge-to-disk luminosity ratios traditionally used in the morphology classifications. For this reason the varying slope of the TF relation according to $\sigma/V_{\rm C}$ corresponds to that seen for varying morphology types classified according to luminosity ratios. 

The parameter $\sigma/V_{\rm C}$ has the additional advantage that it is well defined in all galaxies, and therefore it can be used to expand the present analysis to S0 and elliptical galaxies, where $M_{\rm bulge}/M_{\rm disk}>1$. A future work, based on a larger observed galaxy sample that includes ellipticals and S0s, will address the determination of a generalised galaxy manifold, defined by the circular velocity $[V_{\rm C}]$, the ratio of random-over-total motions$ \sigma/V_{\rm C}$, and the luminosity $L$ (or alternatively the total mass-to-light ratio). 

\bigskip

In this work we characterised the morphology dependence of the Tully-Fisher relation with a physical parameter, and employed it along with circular velocity and luminosity to define a three-dimensional manifold that determines the structure of spiral galaxies. 
We built and analysed a sample of observed galaxies and compared the observed Tully-Fisher relation and the central galaxy velocity dispersion with the predictions by a hierarchical semi-analytic model based on Croton et al. (2006). Our results are the following: 

$\bullet$ the model predicted K-band TF relation is a good match to the data; the hierarchical galaxy formation model fully captures the velocity-luminosity relation for spirals. The model galaxies, classified as Sa, Sb and Sc galaxies with a photometric criterion, show a differentiation of the TF slope, zero-point and scatter with the galaxy type;

$\bullet$ we define a theoretical galaxy velocity dispersion as the component of the rotation curve generated by the spherical, pressure supported mass components, i.e. bulge and dark matter halo, and traced by the stars in the bulge; to compare it with the central (aperture-defined) velocity dispersion measured from galaxy spectra, we compute the mass-average of such component over its density profile;

$\bullet$ we compare the observed ratio of the velocity dispersion over total circular velocity $\sigma/V_{\rm C}$ as a function of $V_{\rm C}$, with the predictions of the semi-analytic model, finding a good match. The model predicts a correspondence between $\sigma/V_{\rm C}$ a and the photometrically-determined galaxy type, with the earlier-types exhibiting a higher $\sigma/V_{\rm C}$ and a larger scatter;

$\bullet$ we divide the observed galaxies in 3 subsamples of different average $\sigma/V_{\rm C}$ following the model trend, and recalculate the TF relation separately for the 3 subsamples; we find that they follow 3 distinct TF relation, with decreasing slope for increasing $\sigma/V_{\rm C}$. The slope of the TF relation for each class of galaxies characterised by $\sigma/V_{\rm C}$ is in agreement with previous results in the literature for Sa, Sb and Sc galaxies.
In addition, this method naturally exclude the TF outliers, thus reducing the scatter on the TF relation;

$\bullet$ we find that $\sigma/V_{\rm C}$ is a good dynamical parameter to characterise galaxy morphology, yielding a classification consistent with the photometrically defined Sa, Sb and Sc types.

We conclude that $\sigma/V_{\rm C}$ is a good, physically motivated third parameter to characterise the TF across the spiral galaxy population. Along with the total velocity $V_{\rm C}$ and the luminosity $M_{\rm K}$, $\sigma/V_{\rm C}$ it thus defines a three-dimensional spiral galaxy manifold that fully characterise the spiral galaxy population.

\section*{Acknowledgments}

We would like to thank the anonymous Referee for her/his comments and suggestions, very beneficial to this work.
We would like to thank Barbara Catinella, Simon Mutch and Darren Croton for their insight and the useful discussions. 
JM is funded by the Australian Research Council Discovery Projects. 
This publication makes use of data products from the Two Micron All Sky Survey, which is a joint project of the University of Massachusetts and the Infrared Processing and Analysis Center/California Institute of Technology, funded by the National Aeronautics and Space Administration and the National Science Foundation.
This research has made use of the NASA/ IPAC Infrared Science Archive, which is operated by the Jet Propulsion Laboratory, California Institute of Technology, under contract with the National Aeronautics and Space Administration.
Funding for SDSS-III has been provided by the Alfred P. Sloan Foundation, the Participating Institutions, the National Science Foundation, and the U.S. Department of Energy Office of Science. The SDSS-III web site is http://www.sdss3.org/.
SDSS-III is managed by the Astrophysical Research Consortium for the Participating Institutions of the SDSS-III Collaboration including the University of Arizona, the Brazilian Participation Group, Brookhaven National Laboratory, Carnegie Mellon University, University of Florida, the French Participation Group, the German Participation Group, Harvard University, the Instituto de Astrofisica de Canarias, the Michigan State/Notre Dame/JINA Participation Group, Johns Hopkins University, Lawrence Berkeley National Laboratory, Max Planck Institute for Astrophysics, Max Planck Institute for Extraterrestrial Physics, New Mexico State University, New York University, Ohio State University, Pennsylvania State University, University of Portsmouth, Princeton University, the Spanish Participation Group, University of Tokyo, University of Utah, Vanderbilt University, University of Virginia, University of Washington, and Yale University.

\end{document}